# Tuning Multipolar Mie Scattering of Particles on a Dielectric-Covered Mirror

Kan Yao[1,2,†,*], Jie Fang[2,†], Taizhi Jiang[2,3], Andrew F. Briggs[4], Alec M. Skipper[4], Youngsun Kim[2], Mikhail A. Belkin[5], Brian A. Korgel[2,3], Seth R. Bank[4], and Yuebing Zheng[1,2,*]

[1] *Walker Department of Mechanical Engineering, The University of Texas at Austin, Austin, Texas 78712, USA*

[2] *Texas Materials Institute, The University of Texas at Austin, Austin, Texas 78712, USA*

[3] *McKetta Department of Chemical Engineering, The University of Texas at Austin, Austin, Texas 78712, USA*

[4] *Microelectronics Research Center and Department of Electrical and Computer Engineering, The University of Texas at Austin, 10100 Burnet Rd. Bldg. 160, Austin, Texas 78758, USA*

[5] *Walter Schottky Institute, Technical University of Munich, Garching 85748, Germany*

[†]These authors contributed equally
[*]Corresponding author(s) ✉: kan.yao@austin.utexas.edu, zheng@austin.utexas.edu

## Abstract

Optically resonant particles are key building blocks of many nanophotonic devices such as optical antennas and metasurfaces. Because the functionalities of such devices are largely determined by the optical properties of individual resonators, extending the attainable responses from a given particle is highly desirable. Practically, this is usually achieved by introducing an asymmetric dielectric environment. However, commonly used simple substrates have limited influences on the optical properties of the particles atop. Here, we show that the multipolar scattering of silicon microspheres can be effectively modified by placing the particles on a dielectric-covered mirror, which tunes the coupling between the Mie resonances of microspheres and the standing waves and waveguide modes in the dielectric spacer. This tunability allows selective excitation, enhancement, and suppression of the multipolar resonances and enables scattering at extended wavelengths, providing new opportunities in controlling light-matter interactions for various applications. We further demonstrate with experiments the detection of molecular fingerprints by single-particle mid-infrared spectroscopy, and, with simulations strong optical repulsive forces that could elevate the particles from a substrate.

**Introduction**

Subwavelength optical resonators play a central role in nanophotonics[1,2]. The design of many devices, such as optical antennas[3-6] and metasurfaces[7-10], is based on rational arrangements of resonant nanoparticles. Among the factors that may affect the functionalities of a nanophotonic device, the achievable optical properties of individual elements are of fundamental importance, because they define the ingredients of inter-particle coupling[11], which in turn determines how flexible the overall device response can be engineered for specific applications. At the single-particle level, conventional strategies for diversifying the available optical properties include shaping the particles, mostly made of plasmonic materials, into complex geometries[12] and using different excitation conditions[13-17]. Dielectric particles with a high refractive index offer another possible route. The multipolar Mie resonances supported by high-index particles, which can be electric or magnetic in nature[18,19], markedly enrich the optical responses from single particles[20,21], leading to the recent emergence of Mie-tronics[22]. However, to enable more exotic optical phenomena after integration, it is highly desirable to equip the design toolbox with novel responses beyond the characteristic resonances of simple elements.

Parallel to designing a particle itself, engineering the dielectric environment (the substrate in particular) allows effective modification of the particle's resonant behaviors. In this context, three substrate configurations are commonly used for different purposes. The simplest one is unstructured substrates. Optically thick metallic films are well known for the ability to perform the function of a mirror[23,24]. It has been shown that the scattering by a silicon (Si) nanosphere at the electric dipolar (ED) resonance can be strongly enhanced through interactions with its mirror image[24-26]. The influence of semi-infinite dielectric substrates on scattering has also been studied extensively for different nano-/microparticles[25-29]. The second configuration adopts the so-called nanoparticle-on-mirror geometry[30,31], which introduces a thin spacer between a particle and a mirror, forming a cavity in the gap that strongly localizes light and results in gap-size-dependent shifts of the resonances[32]. In the third scenario, a wavelength-scale-thick spacer is used. Whereas this scheme is used more often in reflective metasurfaces for enhancing light-matter interactions through Fabry-Pérot-type interference[33,34], recent studies have also demonstrated that it can be used for modification of light scattering from individual nanowires[35,36] and nanoparticles[37,38]. These methods, despite extending the responses of free-standing particles, all rely on interference in the vertical direction and have thus far been mainly applied to the dipolar modes. In pursuit of further extensions to controlling higher-order resonances or more complex mode hybridizations, new mechanisms need to be explored.

In this Article, we show that the multipolar Mie scattering of a Si microsphere (SiMS) can be effectively modified by placing it on different high-index-dielectric-covered mirrors. The role of high-index spacers here is two-fold. In addition to forming a Fabry-Pérot cavity in the vertical direction, it serves as a planar waveguide known as a dielectric-covered ground plane[39]. This brings waveguide modes into the system, which couple with the Mie resonances of the particle in the horizontal directions, giving rise to a new pathway to create complex optical responses. Using high-quality SiMSs and gold mirrors covered by amorphous Si (a-Si) of different thicknesses, we achieve selective excitation, enhancement, and suppression of multipolar Mie resonances in the mid-infrared (MIR, range of wavelengths: ~2-20 μm) regime and waveguide-mode-mediated scattering, extending the optical responses available from individual resonators made of plasmonic or dielectric materials. Potential applications in sensing based on single-particle spectroscopy and in generating repulsive optical forces are demonstrated by experiments and simulations, respectively. We envision that the proposed technique can be generalized to devices containing multiple scatterers or a lattice of resonators[40,41], creating new opportunities in nanophotonics and light-matter interactions.

**Results**

*Scattering properties of Si microspheres on different substrates*

Before investigating the complete system of SiMSs on a dielectric-covered mirror, we first characterize the SiMSs as isolated resonators. Micrometer-sized Si colloids have been studied by Meseguer and coworkers for applications in the ultraviolet, visible, as well as near-infrared regimes[18, 42-45]. Here, we focus mainly on the MIR region, which is of practical interest[46] and less affected by material dispersion and losses. SiMSs sized ~1.5 μm were synthesized by a chemical means with excellent controls over sphericity and size distribution[47-49] (see Materials and Methods and Supplementary Fig. S1). The high quality of the particles was confirmed not only by scanning electron microscopy (SEM) but also by Fourier transform infrared (FTIR) spectroscopy. Figure 1a shows the measured transmittance of a single SiMS (see Materials and Methods and Supplementary Information, section 1), agreeing nearly perfectly with the prediction by Mie theory. Apart from the two minor dents at ~4.25 and 4.8 μm wavelengths (i.e. ~2350 and 2080 $cm^{-1}$ wavenumbers) that correspond to the absorption by carbon dioxide ($CO_2$) in air and Si-H bonds in the SiMS, respectively, every spectral feature is associated with a multipolar Mie resonance (Supplementary Information, section 2 and Figs. S4 and S5).

The scattering properties in free space can be modified by placing the SiMS on a reflective substrate, as illustrated in Fig. 1b. Whereas previous studies observed a strong enhancement of scattering from the ED mode and attributed it qualitatively to an effective magnetic dipole[24] (MD) and the retardation effect[26], we consider another model based on standing waves[36]. Owing to their unique amplitude profile in space, standing waves provide an easy way to control the interactions with matter by positioning the latter at different sites, especially the nodes[50,51] (magnetic field maxima) and anti-nodes[52,53] (electric field maxima). Mirrors are the simplest components to create standing waves. Figure 2a compares the simulated scattering cross sections of the same SiMS sitting directly on two ideal mirrors made of perfect electric (PEC) and magnetic conductors (PMC), respectively. In practice, at MIR wavelengths, a PEC or electric mirror can be resembled by a gold film, while the construction of a PMC or magnetic mirror requires structured surfaces[54]. Although the particle center is not aligned with any node or anti-node of the resulting standing waves, drastically different scattering behaviors are obtained. In contrast to an electric mirror[24], a magnetic mirror enhances the MD mode dramatically and suppresses the ED mode completely. This selectivity is then verified by eigenmode analysis (Supplementary Figs. S9 and S10), finding no solution between the eigenfrequencies of MD and magnetic quadrupole (MQ). Likewise, the electric quadrupole (EQ) and magnetic octupole (MO) are strongly enhanced (suppressed) by the magnetic (electric) mirror. Here, the multipoles are identified by comparing the field distributions with the mode profiles obtained for a freestanding SiMS (Supplementary Fig. S5). An alternative, quantitative approach is multipole expansion[55,56], which has been generalized for the analysis of resonators above a highly reflective substrate based on a different choice of origin[57,58]. Although the present scattering problem is solvable using generalized Mie theory[38,59], the differences in dipolar modes can be explained by a simpler model of two-element arrays in antenna theory[60]. As sketched in Fig. 2a, at normal incidence, the ED and MD modes of a SiMS and their images behind electric and magnetic mirrors have opposite symmetries, corresponding to a phase difference of 0 or $\pi$. Moreover, if neglecting the near-field coupling, the effective distance between the two dipoles approximates a fitting array factor, which accounts for the retardation effect and modifies the radiation pattern of a single dipole (Supplementary Information, sec. 3 and Fig. S11).

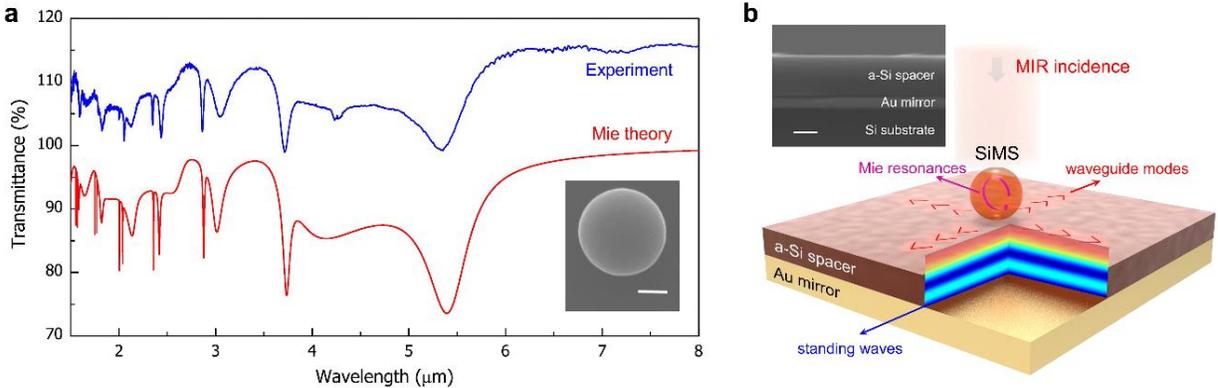

**Figure 1.** Scattering property of a free-standing microsphere and schematic of the system under study. (a) Transmittance spectra of a 1.51 µm SiMS obtained by FTIR spectroscopy (blue curve) and by Mie theory (red curve). The measured spectrum is shifted upwards from the original position by 15% for clarity. Inset: Top-view SEM image of a SiMS sized ~1.5 µm. Scale bar: 500 nm. (b) A SiMS atop an a-Si spacer on a gold mirror. Upon the illumination of MIR light, multipolar Mie resonances of the SiMS and waveguide modes of the substrate are excited and interact differently depending on the particle's location in the standing wave. The simulated standing wave profile is for a discrete case of 3.1 µm wavelength and 600-nm-thick spacer. Inset: Cross-sectional SEM image of an a-Si-covered gold mirror. Scale bar: 200 nm.

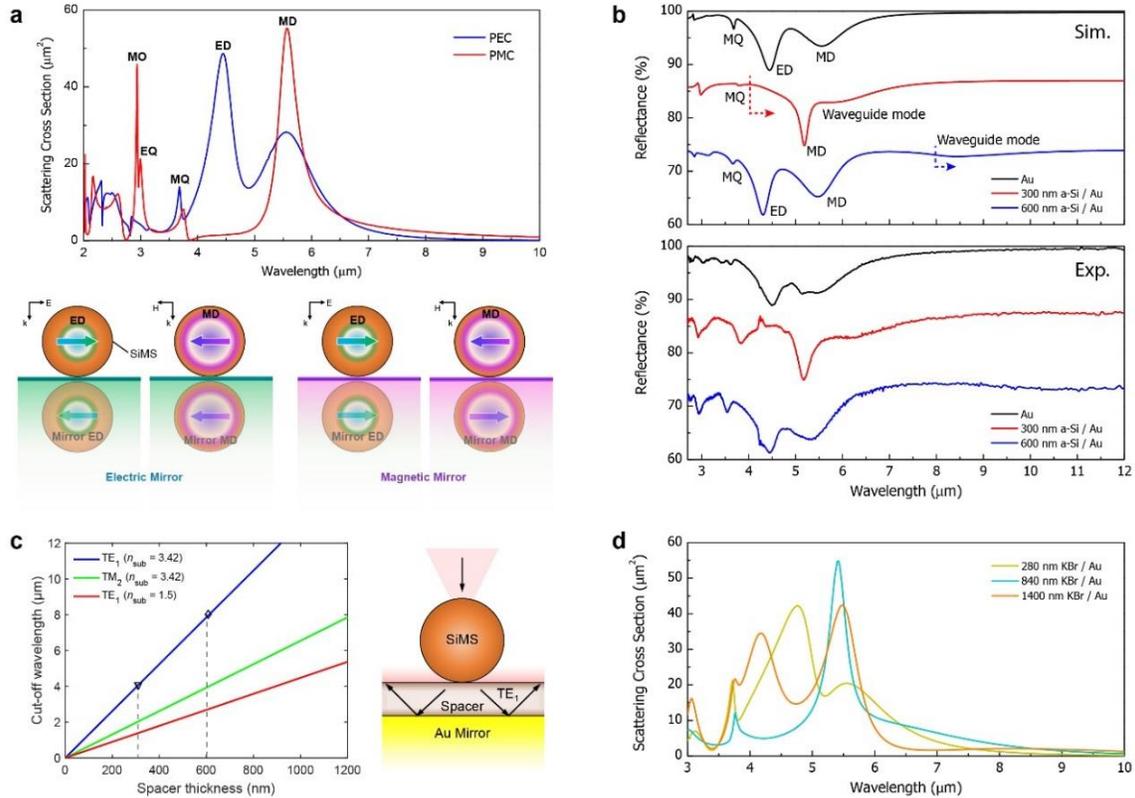

**Figure 2.** Modification of scattering properties by using different substrates. (a) Scattering properties of a SiMS on a perfect electric/magnetic conductor (PEC/PMC). Top: Simulated scattering cross-sections of a 1.51 μm SiMS on PEC (blue curve) and PMC (red curve) at normal incidence exhibit distinct line shapes. Multipolar (especially dipolar) resonances are selectively enhanced or suppressed, depending on the natures of the resonances and of the mirror. Bottom: Schematic of mirror images of a horizontal ED/MD mode produced by an electric mirror (left) and a magnetic mirror (right). (b) Simulated (upper panel) and measured (lower panel) reflectance spectra of individual SiMSs ($a$ = ~1.51 μm) on a gold substrate with no spacer (black curves), a 300 nm a-Si spacer (red curves), and a 600 nm a-Si spacer (blue curves). Spectra are vertically shifted for clarity. The onsets of $TE_1$-mode-mediated scattering features are marked by the vertical dashed lines, and other major spectral features result from Mie resonances. In the measured spectra, features at ~4.25 and 4.8 μm are from the absorption of $CO_2$ in air and Si-H bonds in the SiMS, respectively. (c) Cut-off wavelengths of the low-order waveguide modes supported by a dielectric-covered PEC, as a function of spacer thickness. Symbols on the blue line denote the two spacer thicknesses in (b). Inset: Light scattered off a SiMS is partly coupled to the waveguide modes. (d) Simulated scattering cross-sections of a 1.51 μm SiMS on a gold substrate covered by low-index spacers made of potassium bromide (KBr).

The location of the SiMS (and its mirror image) in a standing wave can be adjusted by adding a transparent spacer. This essentially forms a structure which has the particle on a Gires-Tournois etalon[61]. Given the long wavelengths of MIR light, a-Si that has a high refractive index was used in our study to reduce the spacer thickness. Figure 2b shows the reflectance spectra of a SiMS on three substrates: a bare gold mirror, and two mirrors covered by 300 and 600 nm a-Si, respectively. In addition to good agreement between the simulated and measured spectra, it is noticeable that the line shape of scattering has a strong dependence on spacer thickness. In particular, the ED mode vanishes for the 300-nm-thick spacer, while the MD mode is enhanced as on a magnetic mirror. This selective excitation of MD suggests the utility of the proposed structures as a new type of resonators, supplementary to plasmonic particles supporting only ED and high-index spheres always having ED and MD simultaneously. For the 600-nm-thick spacer, the dipolar modes are restored. All the substrates change the quality factor of the multipoles differently (Supplementary Figs. S5, S9, S10, S12, and S13). Another new feature unseen in the bare mirror cases is the broadband reflection dip. We attribute it to the waveguide modes in the spacer. High-index dielectric slabs support waveguide modes, part of which, as guided waves, cannot be excited directly by plane wave illumination, but are accessible through near-field effects such as scattering[62,63]. With a gold mirror underneath, a spacer acts approximately as a dielectric-covered ground plane, and the cutoff frequency $f_c$ of the $m$ mode is[39] (Supplementary Information, sec. 4)

$$f_c = \frac{m \cdot c_0}{4h \cdot \sqrt{\varepsilon - 1}}. \qquad m = 0, 1, 2, \ldots. \qquad (1)$$

Here, $c_0$ is the speed of light in a vacuum, $h$ and $\varepsilon$ are the thickness and relative permittivity of the spacer, respectively, and $m$ takes odd/even numbers for the transverse electric/magnetic (TE/TM) modes. Except the $TM_0$ mode with a zero cutoff frequency, a guided mode turns into a radiation mode once reaching its cutoff wavelength but retains the role as an extra scattering channel. Figure 2c plots the cutoff wavelengths of $TE_1$ and $TM_2$ modes as functions of spacer thickness. Fair agreement is found between the cutoff of $TE_1$ mode and the onset of the broad features in Fig. 2b, while the influence of higher-order modes is not explicit because of their total spectral overlap with the Mie resonances (Supplementary Figs. S14-S16). Similar modifications can be obtained using low-index spacers with larger thicknesses (Fig. 2d), albeit the waveguide modes are less pronounced. We remark that the effective excitation of a waveguide mode by a Mie resonance requires some spectral overlap[64], which for the current particle size is more evident when the spacer is thinner than 450 nm (Supplementary Fig. S14). As the spacer thickness increases, the cutoff frequencies are red shifted, eventually only intersecting the tail of MD. This leads to a shallow valley in the reflectance (Fig. 2b, blue curves), which nonetheless can be enhanced by using a larger particle or arranging the scatterers into a lattice[65]. Coupling the scattered light into waveguide modes enables a new dimension for tailoring the optical properties of resonators. Compared to the standing waves that only affect the characteristic multipolar resonances of a SiMS, a waveguide mode not only hybridizes with the resonances above its cutoff frequency, but can produce scattering features at extended wavelengths longer than that of the MD. Spatially, it also yields new intensity and polarization profiles capable of fueling light-matter interactions at the surface of the substrate[66]. Fully understanding the interactions between the two sets of modes in the scatterer and in the spacer, nonetheless, requires either a rigorous modal analysis which can describe each channel of the coupling, or, different implementations of mode expansion that allow separation of the multipolar[25] and guided-mode[67] contributions from the total scattering.

*Single-particle spectroscopy for studying light-matter interactions*

One important application of infrared spectroscopy is the identification of chemical substances through their vibrational fingerprints. Whereas techniques like attenuated total reflection and near-field probing are viable options[68], far-field measurements own advantages of simple apparatuses and operations. At the device level, metasurface-based sensing platforms comprising plasmonic[69-74] or dielectric[75] high-$Q$ resonators have been demonstrated, which are optimized for a certain interval of MIR wavelengths, typically covering 5-8 μm (or 1250-2000 cm$^{-1}$). As the multipolar Mie scattering of SiMSs occurs over a broader wavelength range and can be tailored by using different substrates, it is meaningful to further study how these properties will affect light-matter interactions at the single-particle level. Constructive results hold promise for miniaturization[76] of sensing devices or new design principles of metasurfaces.

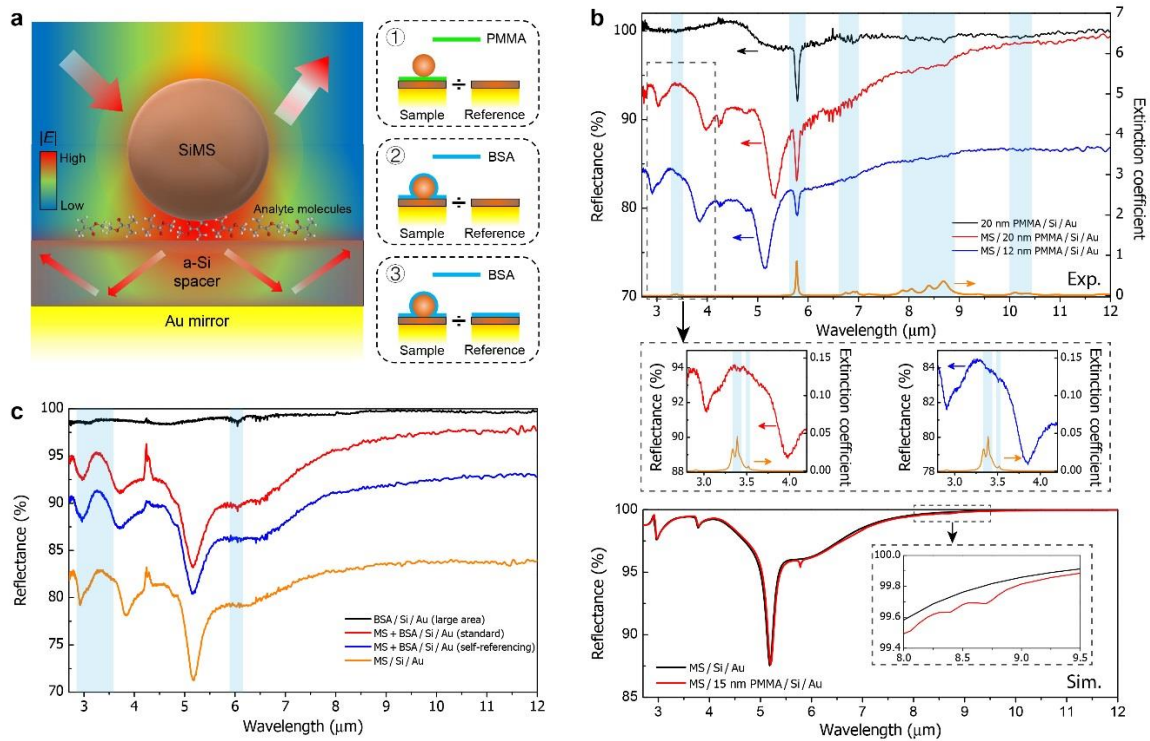

**Figure 3.** Single-particle spectroscopy for sensing. (a) Schematic of the concept. Depending on whether the analyte thin film is deposited before or after drop-casting particles on the substrate, standard or self-referencing measurements are conducted. The standard method uses a bare reference from a different substrate (schemes 1 and 2), whereas the self-referencing method uses a particle-free reference next to the sample area (scheme 3). (b) Detection of PMMA thin films with standard measurements. Top: Measured reflectance spectra of PMMA thin films on a gold substrate covered by a 300 nm a-Si spacer without (black curve) and with (red and blue curves) a SiMS. The former was measured over an area comparable in size to that used in single-particle spectroscopy. Vibrational fingerprints of PMMA are indicated by the shaded strips, in reference to the extinction coefficient of PMMA (orange curve). Without any particles, the desired spectral features are superimposed on an uneven baseline because of the spacer thickness variation between sample and reference areas. In the presence of a SiMS, the same features are superimposed on a better-defined baseline resulting from scattering, and weak signals (e.g. those near 3.4 μm) can be enhanced by the multipolar Mie resonances. Bottom: Simulated reflectance of a 1.51 μm SiMS on a gold

substrate with a 300 nm a-Si spacer and a 15 nm PMMA thin film (red curve), in comparison to the baseline without PMMA (black curve). Inset: Zoom-in of the wavelength interval between 8 and 9.5 μm. (c) Detection of BSA. Measured reflectance spectra of a BSA thin film on a gold substrate covered by a 300 nm a-Si spacer without (black curve) and with (red and blue curves) a SiMS. The former was measured over a large area to get readable signals. Also shown is the reflection spectrum of a SiMS of a similar size on the same substrate but in the absence of BSA (orange curve). Vibrational fingerprints of BSA are marked by the shaded strips. In standard measurements (red curve), both regions around 3 and 6 μm contain noticeable changes in line shape. In the self-referencing case (blue curve), only the broadening near 3 μm was observed. In (b) and (c), measured spectra are vertically shifted for clarity.

For any thin films of analytes, transmission measurements usually suffer from weak signals because of the mismatch between MIR wavelengths and film thicknesses (Supplementary Fig. S17). In theory, depositing the analyte molecules on a dielectric-covered mirror so that they are aligned with an anti-node of the standing wave and measuring reflection is a possible solution, which enhances the absorption from molecules by a factor of up to 4. However, a spacer of thickness $h$ and index $n$ only fulfills this condition over a limited wavelength range around $4nh$, which does not necessarily cover all the molecular vibration bands of interest. At wavelengths far away from $4nh$, the molecule layer may even overlap a node of the standing wave, resulting in weakened absorption. More importantly, in practice, reflection measurements are very sensitive to the spacer thickness. For a-Si, small thickness variations in the order of 5% between the sample and reference areas will suffice to cause uneven baselines containing random features, hindering the reading of meaningful signals. SiMSs on a dielectric-covered mirror have the potential to alleviate the above problems. Figure 3a illustrates three possible schemes for detecting molecular thin films through measuring the scattering of single SiMSs. Two types of molecules, poly(methyl methacrylate) (PMMA) and bovine serum albumin (BSA), are considered to account for different sample configurations (see Materials and Methods for preparation procedures). In the first case, the sample area contains a sub-20-nm-thick PMMA thin film sandwiched between a SiMS and a gold mirror covered by a 300-nm-thick a-Si spacer, while the reference only has the last component, i.e., a bare a-Si-covered mirror. The measured reflectance spectra of the two samples are presented in Fig. 3b (red and blue curves, upper panel), showing readable dips in three out of five vibration bands of PMMA and superimposed on the scattering spectra of SiMSs. Compared with the result from standard reflection measurements in the absence of SiMSs (black curve, also see Materials and Methods and Supplementary Figs. S18 and S19), the scattering by SiMSs produces a well-defined baseline, which is dominant over the random features from the variation of spacer thicknesses. This helps the identification of vibrational fingerprints, especially when the analyte is unknown. Meanwhile, the weak absorption band at 3.3-3.4 μm (2952-2996 cm$^{-1}$, by (O)CH$_3$ stretching[77]) is more perceptible, most likely because of the enhanced local fields from Mie resonances. Similar improvement is derived for the band at 7.9-8.75 μm (1145-1265 cm$^{-1}$, by CH$_2$ bending, C-O-C bending, and C-C-O stretching[68]; see Supplementary Fig. S20), benefiting otherwise from the waveguide mode. Therefore, the modified scattering of SiMSs reduces the influence of spacer nonuniformity and strengthens light-matter interactions at wavelengths where standing waves are inadequate. Further improvements can be expected if multiple particles are arranged into an array. The two groups of dips at 5.78 μm (1730 cm$^{-1}$, by C=O stretching) and 7.9-8.75 μm corresponding to the most prominent absorption bands of PMMA were reproduced by simulations (Fig. 3b, lower panel). For thicker spacers, the influence of spacer nonuniformity on the baseline is more severe, but upon proper overlapping, the multipolar resonances still provide improved capability in detecting certain vibration bands (Supplementary Fig. S21).

Alternatively, molecular thin films can be prepared to cover the entire substrate, including SiMSs. We tested this configuration with BSA using two measurement schemes (Fig. 3a). When the sample signal is referenced to the reflectance from a bare substrate as in the standard procedure, an absorption peak at ~6 μm and broadening of the multipolar resonance near 3 μm are observed (Fig. 3c, red curve), which coincide with the amide I band around 1650 cm$^{-1}$ and a N-H stretching band at 3300 cm$^{-1}$, respectively[78]. Moreover, with conformally coated SiMSs, it becomes possible to perform reference measurements in areas next to the particles, allowing total elimination of any disturbing background signals from another substrate. A reflectance spectrum based on this "self-referencing" scheme is shown in Fig. 3c (blue curve), where the dip at 6 μm is hardly recognizable, but the broadening near 3 μm remains clear. Despite the partial success of detection, this result suggests the limitation of self-referenced measurements. Because the sample and reference areas are coated by nearly the same amount of analyte molecules, standing waves have limited contributions to the signal, and only the absorption bands spectrally overlapping with the modified Mie resonances are detectable. Therefore, to resolve certain vibrational fingerprints with the self-referencing scheme, the scattering baseline needs to be tailored more carefully. This, in principle, can be accomplished by choosing proper particle size and spacer thickness.

*Repulsive optical forces*

Light scattering is naturally accompanied by optical forces. Thus, the modification of multipolar Mie scattering of SiMSs implies new opportunities for optical manipulation. The present system is particularly suitable for studying photonic binding between resonant particles[79-81], because the SiMS not only supports rich resonances, but also has a perfect mirror image that is identical in size. For example, the repulsive forces discussed in ref. 81 may now be realized with a SiMS on a dielectric-covered mirror and by launching a waveguide mode as the incident field. As illustrated in Fig. 4a, the interaction between a SiMS and a TM-polarized guided wave is equivalent to that between a vertical dimer with its gap defined by the spacer and a beam polarized along the dimer axis. If the spacer is relatively thin, all the high-order modes are cut off, while the dominant $TM_0$ mode is loosely confined and impinges on the particle from the side as a quasi-plane-wave, exciting multipolar resonances that have magnitudes different from those in the standard Mie theory. Figure 4b shows the calculated scattering cross section of a SiMS on a 50-nm-thick a-Si spacer, along with the vertical component of the optical force exerted on it. Despite that the gap is filled by a high-index dielectric slab, strong repulsive (away from the substrate) and attractive (toward the substrate) forces are obtained at Mie resonances of certain orders (Fig. 4b, lower panels). Note that stronger scattering does not necessarily generate larger forces. This again highlights the unique property of the structure of SiMSs on a dielectric-covered mirror, where the multipolar Mie scattering mediates the guided modes in the horizontal direction and optical effects in the vertical direction. The strengths of scattering and optical forces are tunable by varying the spacer thickness. For the chosen case of 50-nm-thick a-Si, the strongest repulsive force at ~2330 nm wavelength (magnetic hexadecapole) exceeds 25 pN, assuming the incident power density is $10^{-4}$ W/μm$^2$. This strength is orders of magnitude stronger than gravity and has the potential to overcome van der Waals force if some surface chemistry can be done to avoid direct contact between the SiMS and substrate[81]. Successful realization of the system will allow optical elevation of the SiMSs. For a thicker spacer, the waveguide mode is confined more strongly in the dielectric slab, and the particle and its mirror image are separated by a larger distance. These factors reduce the excitation efficiencies of Mie resonances and their interactions, yielding weakened optical forces.

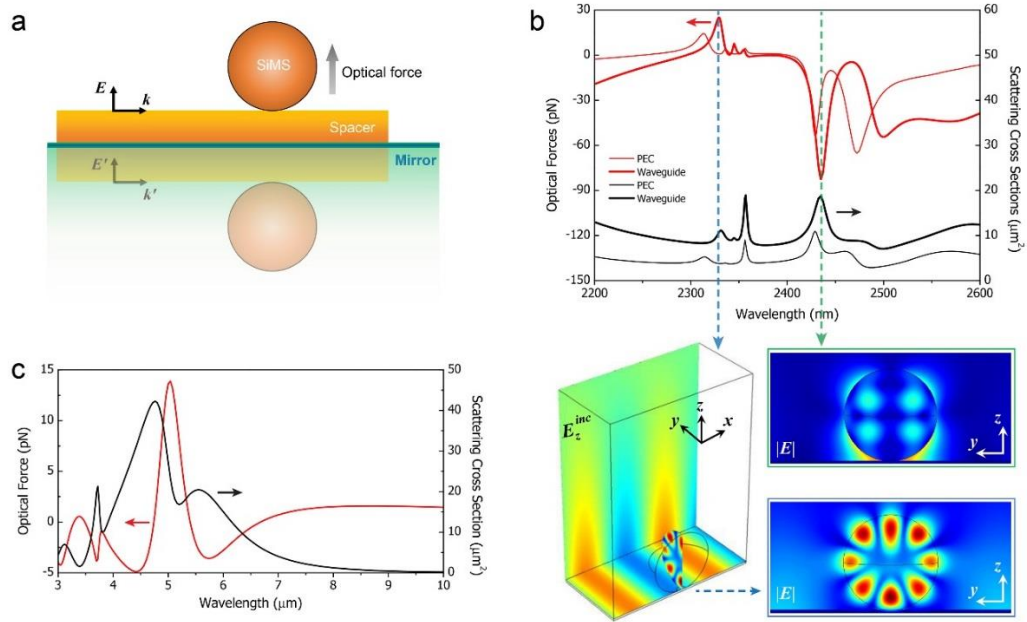

**Figure 4.** Optical force analysis for SiMSs on a dielectric-covered mirror. (a) Schematic of guided-wave-excited repulsive forces, which are analogous to the antibonding forces between two identical SiMSs in a dimer. (b) Simulated optical forces exerted on a 1.51 μm SiMS on a PEC covered by a 50 nm Si spacer (thick red curve), along with the scattering cross-sections (thick black curve). The incident waveguide mode is launched from the side as in (a) and shown in the background of the lower left panel. For comparison, the same spectra (thin curves) are computed for the case where the SiMS is set directly on a PEC and excited by a TM-polarized plane wave at grazing incidence. Electric field distributions at the wavelengths of maximized repulsive and attractive forces are shown in the lower right panels. (c) Same as (b) but for a 1.51 μm SiMS on a KBr-covered PEC, excited by a normally incident plane wave. The spacer thickness is 280 nm.

Lastly, Figure 4c shows the optical force and scattering spectra for the same SiMS on a low-index spacer at normal incidence. Because the optical properties of a SiMS undergo strong modifications as demonstrated above, pushing and pulling forces are expected at the tailored resonances. It is noted that for the case of normal incidence, the force analyses on dipolar modes[23] have recently been extended to higher-order multipoles for suspended particles[82]. However, the present structure still represents a realistic platform for observing photonic binding under different conditions.

**Discussion**

We have shown that the multipolar Mie scattering of SiMSs can be effectively modified by placing them on dielectric-covered mirrors differing in spacer thickness and refractive index. High-index spacers are particularly functional in this system. In addition to generating standing waves in the vertical direction with a compact size, they convert the light scattered off SiMSs into waveguide modes propagating in horizontal directions along the substrate surface and vice versa. These two channels of coupling enable selective excitation, enhancement, and suppression of the multipolar resonances of SiMSs, as well as additional spectral features at longer wavelengths, both of which extend the achievable scattering properties of SiMSs. Further modification is possible by tuning the excitation, such as the illumination angle and polarization. Previous studies have reported angle-dependent quality factors[83] and polarization-controlled excitation of Mie resonances[25,37]. For the present structures, upon illumination at large angles, spectral shifts and variations of scattering intensity are also expected, because oblique incidence effectively changes the position of the particle in the standing wave and the excitation conditions of waveguide modes, including the requirement of in-plane wavevectors and the dependence on polarization. Taking advantage of the extended optical responses, we discuss two potential applications. In the MIR regime, the possibility of detecting molecular thin films based on single-particle spectroscopy is studied. Although the sensitivity is not comparable with what have been achieved by large-area devices such as metasurfaces, our results at the single-particle level are constructive in at least two ways. First, it is possible to design a simple sensing platform using Mie resonators and a mirror covered by a slowly tapered spacer. On this substrate, each resonator has a unique scattering spectrum determined by its own size and local spacer thickness, which can serve as a baseline to identify certain vibrational fingerprints using the self-referencing measurement scheme. Scanning the resonators along the direction of spacer thickness variation provides improved coverage of wavelengths. Second, understanding the responses of individual building blocks benefits the development of multi-element devices. In the present case, the waveguide modes add a new degree of freedom to the design and application of metasurfaces[62]. As the second application and an example of how the scattering channel of waveguide modes can be used reversely for excitation, we numerically demonstrate the generation of strong repulsive optical forces. The proposed structure of Mie resonators on a dielectric-covered mirror could lead to novel photonic devices and opportunities for studying light-matter interactions.

## Materials and Methods

**Silicon microsphere synthesis.** Following our previous work[48,49], we term the amorphous hydrogenated SiMSs as a-SiMS:Hs. A 10 mL titanium batch reactor (High-Pressure Equipment Company (HiP Co.)) was used for the synthesis. First, 126 µL trisilane ($Si_3H_8$, 100%, Voltaix) and 5.5 mL n-hexane (anhydrous, 95%, Sigma-Aldrich) were loaded in the reactor in a nitrogen-filled glovebox. The amount of trisilane determines the particle size, and the amount of n-hexane loaded in the reactor and the reaction temperature is associated with the reaction pressure inside the reactor during the heating process. In this reaction, the temperature is 470 °C, at which the synthesized a-SiMS:Hs have a hydrogen concentration of 5 at.%. An n-hexane amount of 5.5 mL at such temperature ensures a reaction pressure of 34.5 MPa. After adding the reagents, the reactor was sealed by using a wrench inside the glove box. Then a vice was used to tightly seal the reactor after removing it from the glove box. The reactor was heated to the target temperature in a heating block for 10 min to allow the complete decomposition of trisilane. After the reaction, an ice bath was used to cool the reactor to room temperature. Colloidal a-SiMS:Hs were then extracted from the opened reactor. The a-SiMS:Hs were washed by chloroform (99.9%, Sigma-Aldrich) using a centrifuge (at 8000 rpm for 5 min) for three times. The particle diameter is ~1.5 µm.

**Fourier transform infrared spectroscopy.** All measurements, except sensing without SiMSs, were based on single-particle spectroscopy, performed on an FTIR spectrometer (Bruker Vertex 80) combined with a microscope (Hyperion 2000). The area of interest was defined by adjusting a knife-edge aperture on the conjugate image plane. The typical area of the aperture is ~20×20 µm$^2$, which needs to be unchanged for sample and reference measurements. To improve the signal-to-noise ratio, each measurement averaged signals from 2500 scans with a resolution of 4 cm$^{-1}$, which took about 20 min. The knife-edge aperture size was enlarged to at least 150×150 µm$^2$ in the measurements not involving SiMSs, and the number of scans was reduced to about 300 accordingly.

For reflection measurements, a 36× Cassegrain objective (NA = 0.5) was used to focus the incident MIR light on the sample surface and collect the reflected light. For transmission measurements, the same 36× objective was used to collect the transmitted light, and the incident beam was launched by a 15× Cassegrain objective (NA = 0.4) under the sample stage.

**Sample preparation.** The dielectric-covered mirrors were fabricated by sequentially depositing 5-nm-thick titanium, 100-nm-thick gold, 5-nm-thick titanium, and 300/600-nm-thick amorphous silicon on a 4-inch wafer. The last step used low pressure chemical vapor deposition (LPCVD). The wafers were then diced into 1×1 inch$^2$ pieces.

For all the samples, SiMSs were drop-casted on the corresponding substrate. Before drop-casting, the particles stored in chloroform were redispersed in ethanol. For transmission measurements, a transmission electron microscopy (TEM) grid was used as the substrate (Supplementary Fig. S3), of which the thin supporting film serves as a quasi-freestanding platform that minimizes the undesired reflection and absorption by conventional substrates. Compared with the spectra reported in previous works[18,84], the result in Fig. 1a shows superior agreement with Mie theory and substantial improvement in signal-to-noise ratio.

PMMA thin films were prepared by spin coating before SiMSs were drop-casted. To obtain sub-20 nm thicknesses, the 950 PMMA A4 resist was diluted with A thinner at a volume ratio of ~1:5 (for 20-nm-thick films) or ~1:9 (for 12-nm-thick films). After sufficient mixing, the diluted resist was spin coated on the a-Si substrates at 4000 rpm for 45 seconds, followed by baking at 180 °C for 1 min. Film thicknesses were measured using ellipsometry.

To form a thin film of BSA on the sample substrates, i.e., gold mirrors covered by an a-Si spacer, we referred to the procedure described in ref. 85. A 15 mg/mL solution of BSA was prepared by dissolving the required amount of BSA (Fraction V, 97%, Alfa Aesar) in phosphate buffered saline (pH ~7.4, Sigma-Aldrich). Without any treatment to modify the surface hydrophobicity, each substrate (with SiMSs on it) was immersed in the BSA solution for 24 h, then treated with deionized Milli-Q water for 3 min, and finally dried at room temperature. With the given solution PH and ingredients of ions, the adsorbed BSA would form a sub-10-nm-thick thin film on the surface of the a-Si spacers[85].

**Optical Simulations.** The scattering by a free-standing SiMS in Fig. 1(b) was solved by using the standard Mie theory[86]. Other than this case, full-wave simulations were performed using the wave optics module of COMSOL Multiphysics 5.2, a commercial electromagnetic solver based on finite element method. For the scattering by a SiMS on any substrate, a "background field" was first computed with the incident field and substrate but without the particle, which was then used in the second step of the simulation to excite the full structure. The scattering cross section of the SiMS $\sigma$ was obtained by integrating the power flow of the relative fields over a surface enclosing the particle. Transmittance $T$ and reflectance $R$ were converted from the scattering (or extinction, if PMMA was included) cross section using $T = 1 - \sigma/A$ and $R = 1 - \sigma/A$, respectively, where $A$ is a fitting parameter corresponding to the area of the knife-edge aperture. Linear polarization and normal incidence were used in all the simulations that model FTIR measurements, which allows the application of two symmetry planes so that the models can be simplified to containing only one quarter of the full structures. The same reduction was used in the eigenmode analyses, where valid solutions were picked from all the solver-suggested eigenfrequencies by examining the field distributions. The quality factor is derived from the complex eigenfrequency $f_m = \Omega_m + i \cdot \Gamma_m/2$ through $Q_m = \Omega_m/\Gamma_m$, with $\Omega_m$ and $\Gamma_m$ the resonance frequency and damping rate of mode $m$, respectively[87]. To allow for appropriate meshing, SiMSs were truncated at the bottom by 10 nm, contacting the substrate at a facet. The dielectric functions of Si (for both microspheres and spacers) and PMMA were taken as a constant of 11.7 and from ref. 88, respectively. In the presence of a PMMA thin film, instead of including an actual thin layer, transition boundary condition was applied to the substrate surface. In the force analyses, only one symmetry plane is applicable, which reduces the simulation domain to half of the full structure. The incident waveguide mode was launched and absorbed by two opposite sidewalls perpendicular to the symmetry plane, each using a numeric port for boundary mode analysis. Test runs were performed to ensure the height of the simulation domain was greater than the decay length of the waveguide mode. The input power was set to be equal to the power density of a normally incident plane wave multiplied by the decay length and width of the simulation domain. The scattering problem was solved by following the same two-step procedure as in the study of normal incidence. The only difference is that the background field was the waveguide mode passing through a Si slab on PEC with no particle atop it. The optical force was computed by integrating the Maxwell stress tensor over the surface of the SiMS.